\documentclass[prd,twocolumn,aps,superscriptaddress,showpacs]{revtex4}
\usepackage{amssymb}
\usepackage{amsmath,bm}
\usepackage{graphicx}
\usepackage[normalem]{ulem}
\usepackage[dvips]{color}
\usepackage{subfigure}
\newcommand{\be}{\begin{equation}}

\newcommand{\ee}{\end{equation}}
\newcommand{\bea}{\begin{eqnarray}}
\newcommand{\eea}{\end{eqnarray}}

\renewcommand\sout{\bgroup \color{red} \ULdepth=-.5ex \ULset}

\begin{document}

\title{Quark matter at finite temperature and proto-quark stars with the axion effects in SU(3) Nambu-Jona-Lasinio model}
\author{Peng-Cheng Chu }
\email{kyois@126.com}
\affiliation{The Research Center for Theoretical Physics, Science School, Qingdao University of Technology, Qingdao, 266033, China}
\author{Jiao-Jiao Wang }
\affiliation{The Research Center for Theoretical Physics, Science School, Qingdao University of Technology, Qingdao, 266033, China}
\author{Ke Liu }
\affiliation{The Research Center for Theoretical Physics, Science School, Qingdao University of Technology, Qingdao, 266033, China}
\author{Peng Wu }
\affiliation{The Research Center for Theoretical Physics, Science School, Qingdao University of Technology, Qingdao, 266033, China}
\author{Yu-Heng Liu}
\affiliation{The Research Center for Theoretical Physics, Science School, Qingdao University of Technology, Qingdao, 266033, China}

\author{He Liu }
\email{liuhe@qut.edu.cn}
\affiliation{The Research Center for Theoretical Physics, Science School, Qingdao University of Technology, Qingdao, 266033, China}
\author{Hong-Ming Liu }
\email{liuhongming13@126.com}
\affiliation{The Research Center for Theoretical Physics, Science School, Qingdao University of Technology, Qingdao, 266033, China}
\author{Xiao-Hua Li }
\email{lixiaohuaphysics@126.com}
\affiliation{School of Nuclear Science and Technology, University of South China, Hengyang, 421001, China }
\affiliation{Cooperative Innovation Center for Nuclear Fuel Cycle Technology $\&$ Equipment, University of South China, Hengyang, 421001, China }

\author{Min Ju }
\email{ jumin@upc.edu.cn}
\affiliation{College of Science, China University of Petroleum (East China), Qingdao, 266580, China }

\author{Xu-Hao Wu }
\email{wuhaoysu@ysu.edu.cn}
\affiliation{School of Science, Yanshan University, Qinhuangdao 066004, China}
\author{Xiao-Min Zhang }
\email{zhangxm@mail.bnu.edu.cn}
\affiliation{The Research Center for Theoretical Physics, Science School, Qingdao University of Technology, Qingdao, 266033, China}
\author{Ying Zhou }
\email{yingzhow@163.com}
\affiliation{School of Physical Science and Technology,
Inner Mongolia University, Hohhot, 010021, China}


\begin{abstract}
We investigate the thermodynamical properties of strange quark matter (SQM) and proto-quark stars (PQSs) within the SU(3) Nambu-Jona-Lasinio (NJL) model at finite temperature, specifically incorporating the effects of axion fields and vector interactions. Our results demonstrate that these interactions significantly influence the equation of state (EoS), constituent quark masses, entropy density, and the maximum star mass of PQSs at the isentropic stages along the star evolution line. Furthermore, we reveal a distinct thermodynamic signature in the early evolution: the presence of trapped neutrinos leads to a substantial increase in electron number density while simultaneously suppressing the core temperature compared to the neutrino-free case. These findings may highlight the crucial role of the axion effects, flavor-dependent vector interactions, and particle composition in determining the observable properties of compact stars at finite temperature.
\end{abstract}

\maketitle

\section{Introduction}

As a fundamental probe in nuclear physics and astrophysics, the equation of state (EoS) of strongly interacting matter provides critical insights into the thermodynamic nature of the medium, which plays roles in understanding the phase transition in the quantum chromodynamics (QCD) phase diagram, the intrinsic thermodynamic properties of both nuclear matter and quark matter, as well as the internal structure and observable properties of compact stars \cite{Fukushima11, Brandes24}. While heavy-ion collisions (HICs) allow people to explore strongly interacting matter under the extreme conditions of finite temperature and high baryon density in terrestrial laboratories, astrophysical observations of compact stars (including neutron stars (NSs), quark stars (QSs), and hybrid stars (HSs)) offer an alternative natural laboratory to test the EoS of isospin asymmetric nuclear matter and quark matter~\citep{Lattimer04,Steiner05}. Among these candidates of compact stars, strange quark stars, which are hypothesized to form via the conversion of neutron stars~\citep{Bom04,Sta07,Her11, Lai11,Alford03, Alford05, Baldo03, Ippolito08, de11, Bonanno12,Chu16b,Chu17,Chu23b,Chu20,Liu22,Chu23}, consist of absolutely stable strange quark matter (SQM). This exotic matter comprises deconfined $u$, $d$, and $s$ quarks and leptons in $\beta-$equilibrium condition, characterized by significant isospin asymmetry \cite{Steiner01,SteinerPRL01,Ni06,Burgio08,YA09,Chu19,Chen2012}. Investigating the potential existence of QSs is therefore crucial for unraveling the properties of SQM across a wide range of baryon densities, chemical potentials, and temperatures~\citep{Iva69, Ito70, Bodmer71, Witten84, Far84, Alc86, Web05,Terazawa79,Ste98}. In these years, astrophysical observations have provided the mass and radius range of several heavy pulsars, putting crucial constraints on the phenomenological models of nuclear matter. Notable examples include PSR J1614-2230 in 2010 ($1.97\pm0.04~M_{\odot}$) \cite{Demorest10}, PSR J0348+0432 in 2013 ($2.01\pm0.04~M_{\odot}$) \cite{Ant13}, and PSR J0740+6620 in 2019 (updated to $2.08\pm0.07~M_{\odot}$) \cite{Cromartie19,Fonseca21,Miller21}. Furthermore, recent observations have revealed even more massive candidates: PSR J0952-0607 with $M=2.35\pm0.17~M_{\odot}$ \cite{Romani22}, the secondary component of GW190814 with $2.50~M_{\odot}-2.67~M_{\odot}$ \cite{Abbott2020}, and PSR J0514-4002E with $2.09~M_{\odot}-2.71~M_{\odot}$ \cite{Ewan24}. Describing the heavy pulsars listed above as QSs presents a significant challenge, as the EoS of the quark star matter in most phenomenological models is soft due to the additional strange quark degree in SQM \cite{Gle00,Web99,Cho74,Alf05,Reh96,Han01,Rus04,Men06,Fre77,Fre78,Kur10,Rob94,Zon05,Qin11,Li11,Peng99, Peng00,Pen08}, which decreases the maximum star mass of strange quark stars. In order to provide stiffer EoSs for supporting more massive compact stars, researchers have considered increasing the vector and isovector effects inside the isospin asymmetric star matter (large symmetry energy is employed to stiffen the EoSs)~\cite{Steiner01,SteinerPRL01,Ni06,Burgio08,YA09,Chen2012}, while other studies have focused on modifying the (isospin-) density dependence of the equivalent quark mass to obtain massive QSs \cite{Peng99, Peng00,Pen08,Chu2014,Chu2019,Chu21,Chu21b,Chu2024PRD,Chu2026EPJC}.

The explosion of a Type II supernova occurs when massive stars deplete their nuclear fuel, causing the core to undergo gravitational collapse. The remnant of this gravitational collapse results in either a neutron star or a black hole, depending on the progenitor's initial conditions (specifically the core mass) \cite{Prakash97}. Since the lack of knowledge of the transition from proto-neutron star (PNS) to proto-quark star (PQS) during the complex burning process, previous works investigate the properties of PQSs with SQM at finite temperature at different isentropic stages \cite{Alcock86,Gupta03,Shen05,Dexheimer13,Dexheimer14}. The investigation of PQSs is of importance for understanding the temperature effects on the thermodynamic properties of SQM, the heating process driven by the diffusing neutrinos at the isentropic stages, and the properties of the EoS of strongly interacting matter at finite temperature.

Additionally, the axion, which is predicted as the pseudo-Goldstone boson arising from the spontaneous breaking of Peccei-Quinn symmetry \cite{Peccei77a,Peccei77b,Chatterjee12, Cortona16}, is regarded as a well-motivated dark matter candidate. This hypothesis is actively investigated by experiments such as the Particle and Astrophysical Xenon Experiments (PandaX) \cite{PandaX24}. Furthermore, the presence of axions may significantly influence the maximum mass of compact stars at zero temperature \cite{Karkevandi22, Bruno22,Chu2024PRD}. In this work, we incorporate the effects of axion fields and vector interactions at finite temperature within the SU(3) Nambu-Jona-Lasinio (NJL) model so as to produce stiffer equations of state at finite temperature, thereby investigating the properties of massive PQSs.

The paper is organized as follows. In Sec.II, we derive the theoretical formulism for SQM within the SU(3) NJL model at finite temperature by considering the axion field and vector interactions. Then we present our numerical results about the properties of quark star matter and proto-quark stars at finite temperature isentropic stages in Sec. III. Finally in Sec. IV, we provide our conclusion and discussion.

\section{THE THEORETICAL FORMULISM}
\subsection{The SU(3) NJL model with axion effects}
\label{NJL}
The Lagrangian density within the SU(3) NJL model by considering the axion contribution for SQM is written as
\begin{eqnarray}
\mathcal{L}&=&\bar{\psi}(i\partial\!\!\!\slash-\hat m_c)\psi-K\{e^{i\theta}{\det}_f[\bar{\psi_f} (1+\gamma_5)\psi_f]\notag\\
&+&e^{-i\theta}{\det}_f{[\bar{\psi_f }(1-\gamma_5)\psi_f]}\} \notag\\
&+&G_S\sum\limits _ {a=0}^8 [(\bar{\psi _f}\lambda_a \psi_f)^2 +(\bar{\psi_f}i\gamma_5\lambda_a\psi_f)^2]\notag\\
&-&G_V{\sum\limits_{a=0}^8[(\bar{\psi}\gamma^{\mu}\lambda^a\psi)^2
+(\bar{\psi}i\gamma^\mu\gamma_5\lambda^a\psi)^2]}.
\label{Lag}
\end{eqnarray}
The quark field in the Lagrangian density is defined as $\psi=(u,d,s)^T$, and $\hat{m}_c={\rm diag}(m_u,m_d,m_s)$ represents the mass matrix for the current quark mass. $K$ denotes the coupling constant of the 't Hooft term, which introduces the six-point interaction. The parameter $\theta$ corresponds to the chiral rotation angle generated by the axion field contribution in SQM \cite{Chatterjee12}. $G_S$ is the coupling constant for the scalar interaction, $\lambda_a $ means the Gell-Mann matrices, and $G_V$ is the coupling constant of the flavor-dependent repulsion vector interaction among $u$, $d$, and $s$ quarks. Considering the mean-field approximation, the Lagrangian density for SQM can be derived as

\begin{eqnarray}
\mathcal{L}_{M}&=&\bar{\psi}[\gamma_{\mu}i\partial^ \mu-\hat{M}+\hat{\mu}\gamma_0 ]\psi -2G_S\sum\limits_{i=u,d,s}(\sigma_i^2+\eta_i^2)\notag\\
&+&4K\Big[\cos\theta(\sigma_u\sigma_d\sigma_s-\sigma_s\eta_u\eta_d-\sigma_u\eta_d\eta_s-\sigma_d\eta_u\eta_s)\notag \\
&-&\sin\theta(\eta_u\eta_d\eta_s-\sigma_u\sigma_s\eta_d-\sigma_d\sigma_s\eta_u-\sigma_u\sigma_d\eta_s)\Big]\notag\\
&+&2G_V(n_u^2+n_d^2+n_s^2).
\end{eqnarray}
Here $\hat{\mu}$ represents the chemical potential for quarks, and $
\hat{n}=diag(n_u,n_d,n_s)
$ denotes the number density of each flavor of quarks. The quantities $\sigma_f=\langle\bar{\psi_f}\psi_f\rangle$ and $\eta_f=\langle\bar{\psi_f}i\gamma_5\psi_f\rangle$ are defined as the quark condensates in the scalar and pseudoscalar channels, respectively.

The constituent quark mass of the Lagrangian density with the axion contribution can be written as
\begin{equation}
M_i=\sqrt{{M_i^s}^2+{M_i^p}^2},
\end{equation}
where $M_i^s$ and ${M_i^p}$ stand for the scalar interaction part and pseudoscalar interaction part for the i-th flavor of the constituent quark mass, and the scalar interaction part of the constituent quark mass can be given as
\begin{eqnarray}
M_i^s&=&m_{i0}-4G_S\sigma_i+2K\Big[\cos\theta(\sigma_j\sigma_k-\eta_j\eta_k)\notag\\
&+&\sin\theta(\sigma_j\eta_k+\sigma_k\eta_j)\Big],
\end{eqnarray}
where $(i,j,k)=(u,d,s)$. The scalar contribution and pseudoscalar contribution to the quark condensates can be obtained as
\begin{equation}
\sigma_i=-2N_C\int\Big(1-\frac{1}{1+e^{\beta{(\epsilon_i-\tilde{\mu})}}}-\frac{1}{1+e^{\beta{(\epsilon_i+\tilde{\mu})}}}\Big)
\frac{M_i^s}{\epsilon_i}\frac{\text d^3p}{2\pi^3} \\,
\end{equation}
and
\begin{equation}
\eta_i=2N_C\int\Big(1-\frac{1}{1+e^{\beta(\epsilon_i-\tilde{\mu})}}-\frac{1}{1+e^{\beta(\epsilon_i+\tilde{\mu})}}\Big)
\frac{M_i^p}{\epsilon_i}\frac{\text d^3p}{2\pi^3} \\.
\end{equation}
Here $N_C=3$ , $\epsilon_i=\sqrt{M_i^2+p^2}$, and the effective chemical potential $\tilde\mu$ is written as
\begin{equation}
\tilde{\mu}_{f}=\mu_{f}-4G_{V}n_{f}.
\end{equation}

The pseudoscalar interaction part of the i-th flavor of the quark mass can be obtained as
\begin{equation}
M_i^p=4G_S\eta_i+2K\Big[\cos\theta(\sigma_k\eta_j+\sigma_j\eta_k)-\sin\theta(\sigma_j\sigma_k-\eta_j\eta_k)\Big].
\end{equation}

Then we can further obtain the thermodynamical potential $\Omega_q$ of the quark part
\begin{eqnarray}
\Omega_q&=&\sum\limits_{i=u,d,s}\Big[-i\int\frac{d^4p}{(2\pi)^4}
\text{tr}\ln{\Big\{\frac{1}{T}[p\!\!\!/-\hat{M_f}+\gamma_0\tilde{\mu}_f]}\Big\}\notag \\
&+ & 2G_S(\sigma_i^2+\eta_i^2)\Big]-4K\Big[\cos\theta(\sigma_u\sigma_d\sigma_s-\sigma_s\eta_u\eta_d\notag \\
&-&\sigma_u\eta_d\eta_s-\sigma_d\eta_u\eta_s)-\sin\theta(\eta_u\eta_d\eta_s-\sigma_u\sigma_s\eta_d-\sigma_d\sigma_s\eta_u\notag\\
&-&\sigma_u\sigma_d\eta_s)\Big]-2G_V(n_u^2+n_d^2+n_s^2).
\label{pq}
\end{eqnarray}

We can employ $\mathcal{F}=\Omega+\sum_i \mu_i n_i$ to obtain the free energy density for quarks at finite temperature, and the free energy density of the quark part can be given as
\begin{eqnarray}
\mathcal{F}_q&=&-2N_c \sum_{f=u,d,s}\int\frac{\text{d}^3p}{(2\pi)^3}\Biggr[\epsilon_f-\frac{\epsilon_f}{1+e^{({\epsilon_f-\tilde{\mu}_f})/{T}}}\notag\\
&-&\frac{E_f}{1+e^{({\epsilon_f+\tilde{\mu}_f})/{T}}}\Biggr]+ 2G_S(\sigma_i^2+\eta_i^2)\notag \\
&-&4K\Biggr[\cos\theta(\sigma_u\sigma_d\sigma_s-\sigma_s\eta_u\eta_d-\sigma_u\eta_d\eta_s-\sigma_d\eta_u\eta_s)\notag \\
&-&\sin\theta(\eta_u\eta_d\eta_s-\sigma_u\sigma_s\eta_d-\sigma_d\sigma_s\eta_u-\sigma_u\sigma_d\eta_s)\Biggr]\notag\\
&-&2G_V(n_u^2+n_d^2+n_s^2) \notag\\
&+&\sum_{f=u,d,s}(\mu_f-\tilde\mu_f)n_f-\mathcal{F}_{q0}.
\label{Etotal}
\end{eqnarray}

Then the energy density can be obtained by considering $\mathcal{F}=\mathcal{E}-TS$, where $S=-\frac{\partial \Omega}{\partial T}$ means the entropy density of SQM at finite temperature.
The particle density of quarks and leptons at finite temperature can be obtained as
\begin{eqnarray}
 n_i = \frac{g_i}{2\pi^2}\int\Big[\frac{1}{1+e^{(\epsilon_i-\tilde\mu_i)/T}} -\frac{1}{1+e^{(\epsilon_i+\tilde\mu_i)/T}}\Big]p^2\text{d}p,\notag\\
\end{eqnarray}
where $g_i=6$ for quarks while $g_i=2$ for leptons.
The SQM inside the QSs at finite temperature is composed of three flavors of quarks ($u$, $d$, and $s$)
and leptons ($e^-$, $\mu$, $\nu_{e}$ and $\nu_{\mu}$). The electric charge neutrality can be expressed as
\begin{eqnarray}
\frac{2}{3}n_u=\frac{1}{3}n_d+\frac{1}{3}n_s+n_e+n_{\mu}.
\end{eqnarray}
And the weak $\beta-$equilibrium condition can be obtained as
\begin{eqnarray}
\mu_u+\mu_e-\mu_{\nu_e}=\mu_d=\mu_s, \notag\\
\mu_{\mu}=\mu_{e}~ ~\text{and}~~\mu_{\nu_{\mu}}=\mu_{\nu_{e}}.
\end{eqnarray}

Following the previous studies \cite{Chu17,Chu21b}, we describe the evolution of PQSs using a framework similar to that of PNSs, characterized by different isentropic stages along the star evolution line. At the very beginning of the birth of proto-compact stars after the supernova explosions, the entropy per baryon of the 1st isentropic stage is set as $S/n_B=1$, while the fraction of the lepton number with trapped neutrinos is constrained as 0.4 ($Y_{l}=Y_{e}+Y_{\mu}+Y_{\nu_{l}}=Y_{e}+Y_{\mu}+Y_{\nu_{e}}+Y_{\nu_{u}}=0.4$). Then in the following 10-20 seconds, the neutrinos diffuse and heat the star matter in the 2nd isentropic stage, which futher increases the entropy per baryon. In the final stage, the star continues cooling down and eventually evolves into a cold QS. In this work, we describe the star evolution of PQSs by considering three snapshots as \cite{Dexheimer14,Steiner2000,Reddy90,Menezes15,Shao11}
\begin{eqnarray}
(\mathrm{I})S/n_{B}=1,~ Y_{l}=0.4,\notag\\
(\mathrm{II})S/n_{B}=2,~Y_{\nu_{l}}=0,\notag\\
(\mathrm{III})S/n_{B}=0,~ Y_{\nu_{l}}=0.
\end{eqnarray}

\section{Results and discussions}

In this work, we employ the parameter set of SU(3) NJL model from the Ref. \cite{Pereiraet16,Chu25}, referred to as PCP following the authors' initials in Ref.~\cite{Pereiraet16}, with $\Lambda=630.0$ MeV, $m_{u}=m_{d}=5.5$ MeV, $m_{s}=135.7$ MeV, $G_S\Lambda^2=1.781$, and $K\Lambda^5=9.29$. In Ref. \cite{Chu2024PRD}, the authors discuss the parameter space satisfying the absolute stability condition of SQM within the SU (3) NJL model at zero temperature. This condition implies that the minimum energy per baryon of SQM at zero temperature must be lower than that of the most stable observed nuclei ($M(^{56}\text{Fe})/56=930$ MeV), based on the hypothesis that SQM might be the true ground state of strongly interacting matter \cite{Witten84,Far84}. The results indicate that the minimum energy per baryon at zero temperature becomes smaller than $930$ MeV when $\theta$ is larger than $0.35\pi$ using the PCP parameter set, thereby satisfying the absolute stability condition for SQM within SU(3) NJL model.

\begin{figure}[tbp]
\includegraphics[scale=0.34]{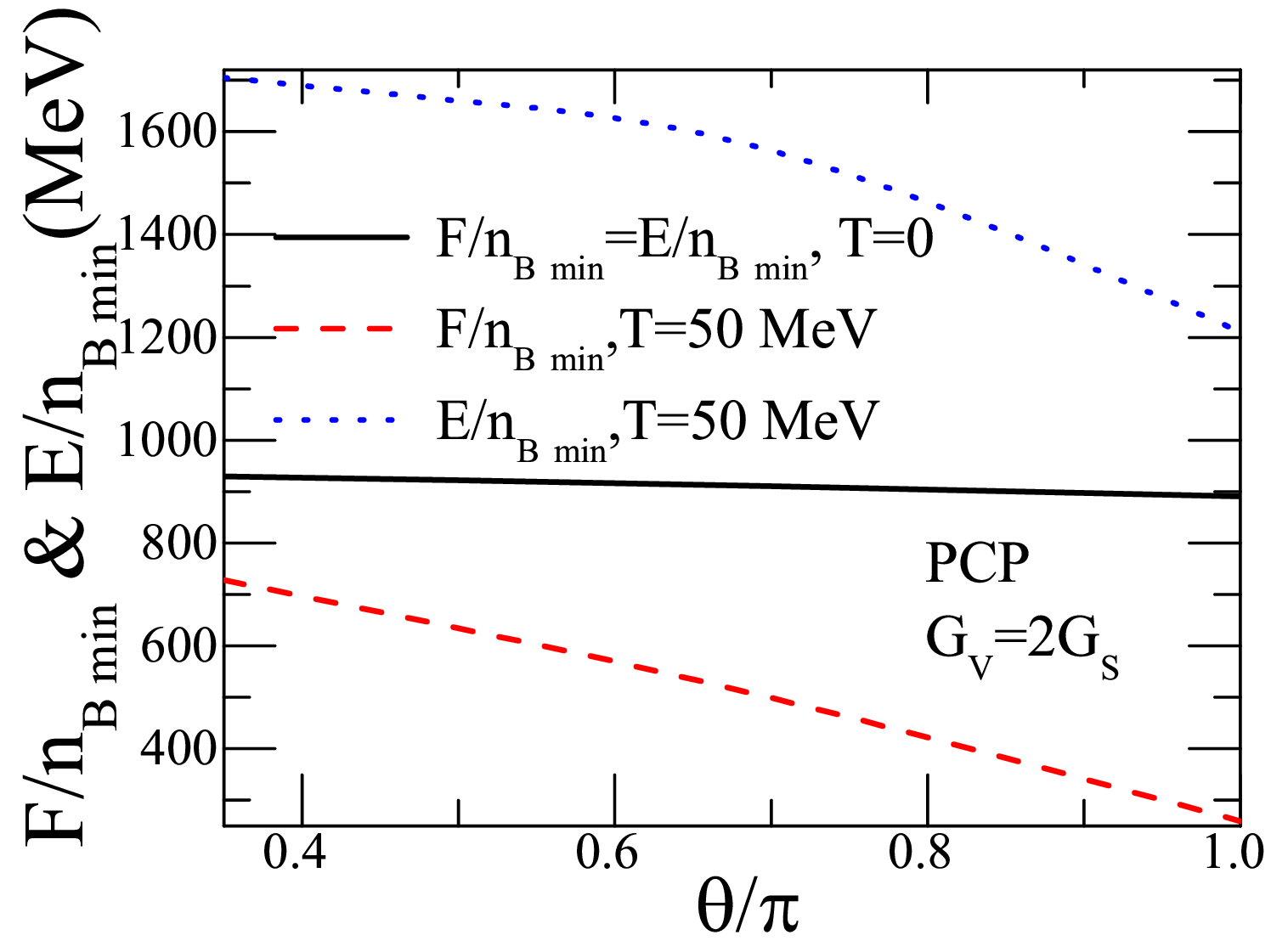}
\caption{(Color online) The minimum value of the free energy per baryon and energy per baryon of SQM as functions of $\theta$ within SU(3) NJL model. }
\label{SEFE}
\end{figure}

In Fig. 1, we present the minimum free energy per baryon and energy per baryon of SQM as functions of $\theta$ at finite temperature ($T = 0$ and $50$ MeV), calculated within SU(3) NJL model using the PCP parameter set. The flavor-dependent vector interaction coupling constant is set as $G_V=2G_S$. This choice allows the model to describe both the black widow pulsar PSR J0952-0607 ( $M=2.35\pm0.17~M_{\odot}$) and the secondary component ($2.50~M_{\odot}-2.67~M_{\odot}$) of GW190814 as QSs at zero temperature, with the maximum star mass reaching $2.51~M_{\odot}$. Fig. 1 demonstrates that both ${F/n_B}_{min}$ and ${E/n_B}_{min}$ generally decrease with increasing $\theta$. Notably, the dependence on $\theta$ is much stronger at finite temperature case $T=50$ MeV than at zero temperature case (the curve for ${F/n_B}_{min}$ at $T=0$ remain relatively flat from $930$ MeV to $891$ MeV when $\theta$ increases from $0.35\pi$ to $\pi$, while the curve for ${F/n_B}_{min}$ at $T=50$ MeV drops significantly from $728$ MeV to $259$ MeV as $\theta$ approaches to $\pi$). This indicates that the stability of SQM is greatly enhanced by the axion field effect, particularly in the finite temperature regime.

\begin{figure}[tbp]
\includegraphics[scale=0.355]{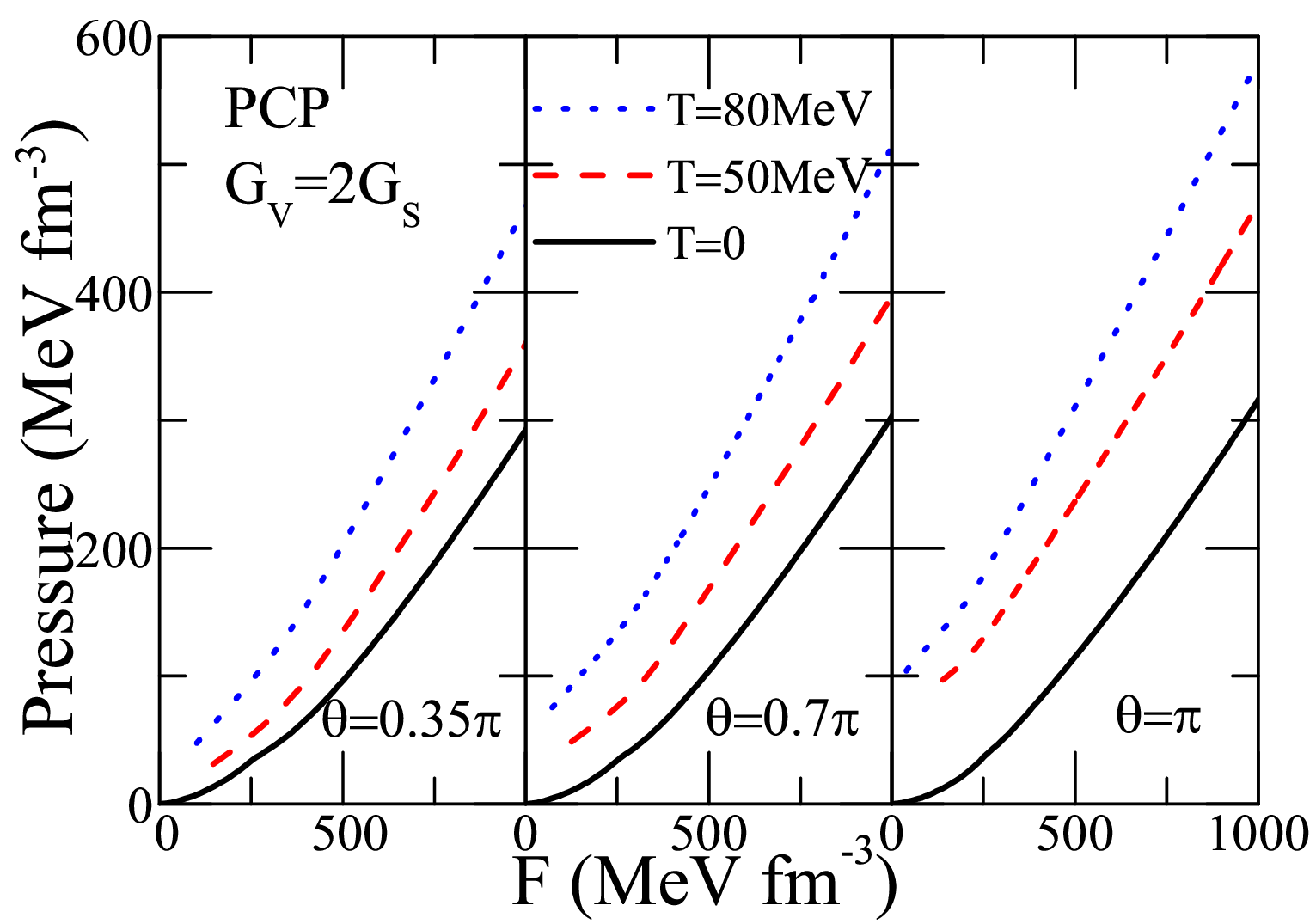}
\caption{(Color online) Pressure as a function of free energy density for SQM at finite temperature with different $\theta$ within SU(3) NJL model. }
\label{MASS}
\end{figure}

As shown in Fig.~2, we calculate the pressure of SQM as a function of the free energy density at finite temperature by using the parameter set PCP with $G_V=2G_S$. It can be seen that for a fixed $\theta$, the pressure increases with temperature, while the pressure also increases when $\theta$ increases from $0.35\pi$ to $\pi$. These results indicate that the EoS of SQM at finite temperature within SU(3) NJL model becomes stiffer with increasing axion effects and temperature.

\begin{figure}[tbp]
\includegraphics[scale=0.35]{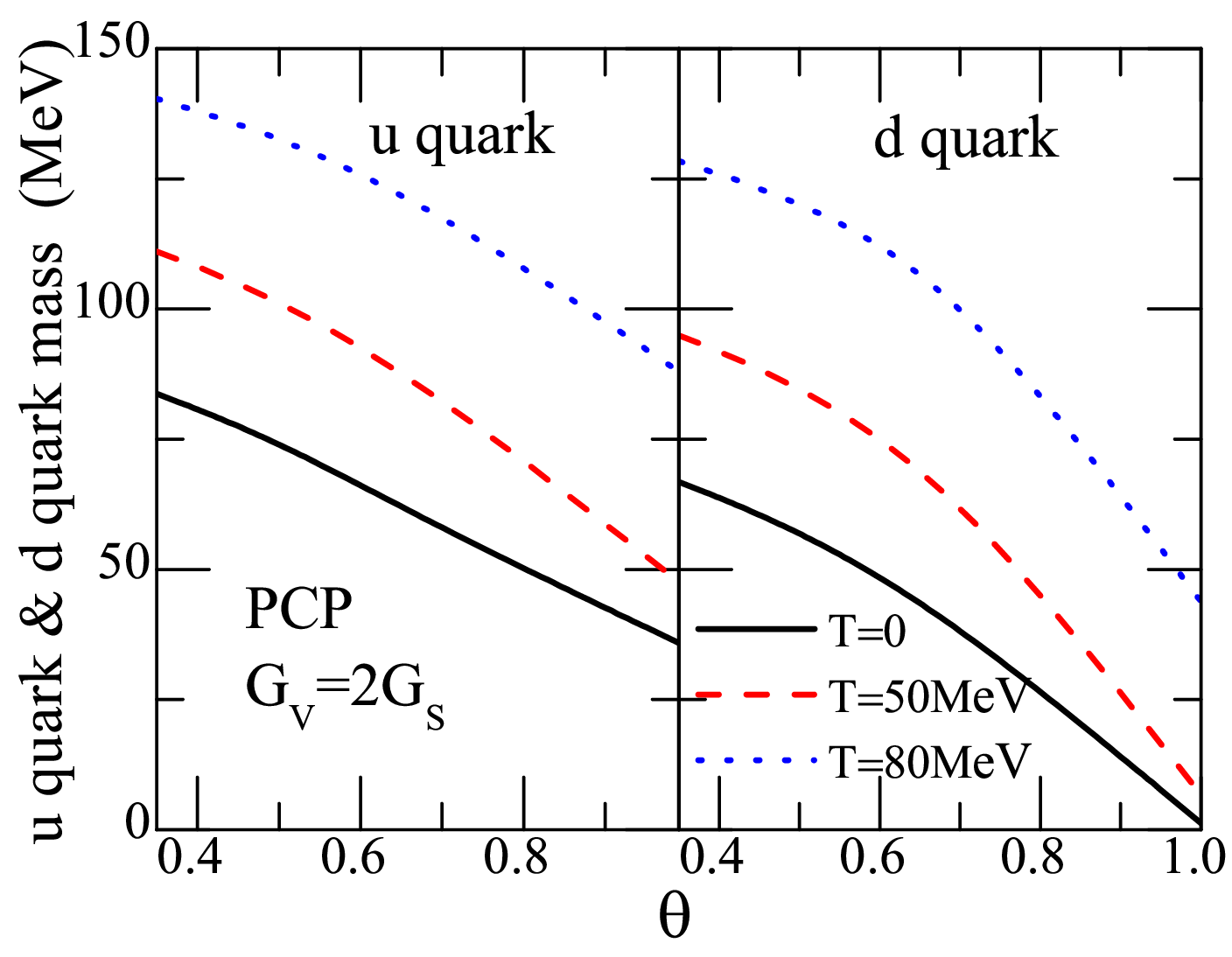}
\caption{(Color online) The constituent quark mass of $u$ and $d$ quarks as functions of $\theta$ when $G_V=2~G_S$ at zero and finite temperature. }
\label{fraction}
\end{figure}

In Fig.3, we show the constituent quark mass of $u$ and $d$ quarks as functions of $\theta$ with the baryon density $n_B$ being fixed at $2~n_0$ ($n_0$ is the saturation density of nuclear matter). One can see from Fig. 3 that the constituent quark mass of $u$ and $d$ quarks for all the cases decreases with $\theta$, for both zero and finite temperature cases. Furthermore, for a fixed $\theta$, the constituent quark mass increases with temperature, with the mass of $d$ quark being smaller than that of $u$ quark (the fraction of $d$ quarks is larger than that of $u$ quarks because of the charge neutrality in $\beta-$equilibrium, which may increase the quark condensate and further decrease the constituent quark mass).

\begin{figure}[tbp]
\includegraphics[scale=0.32]{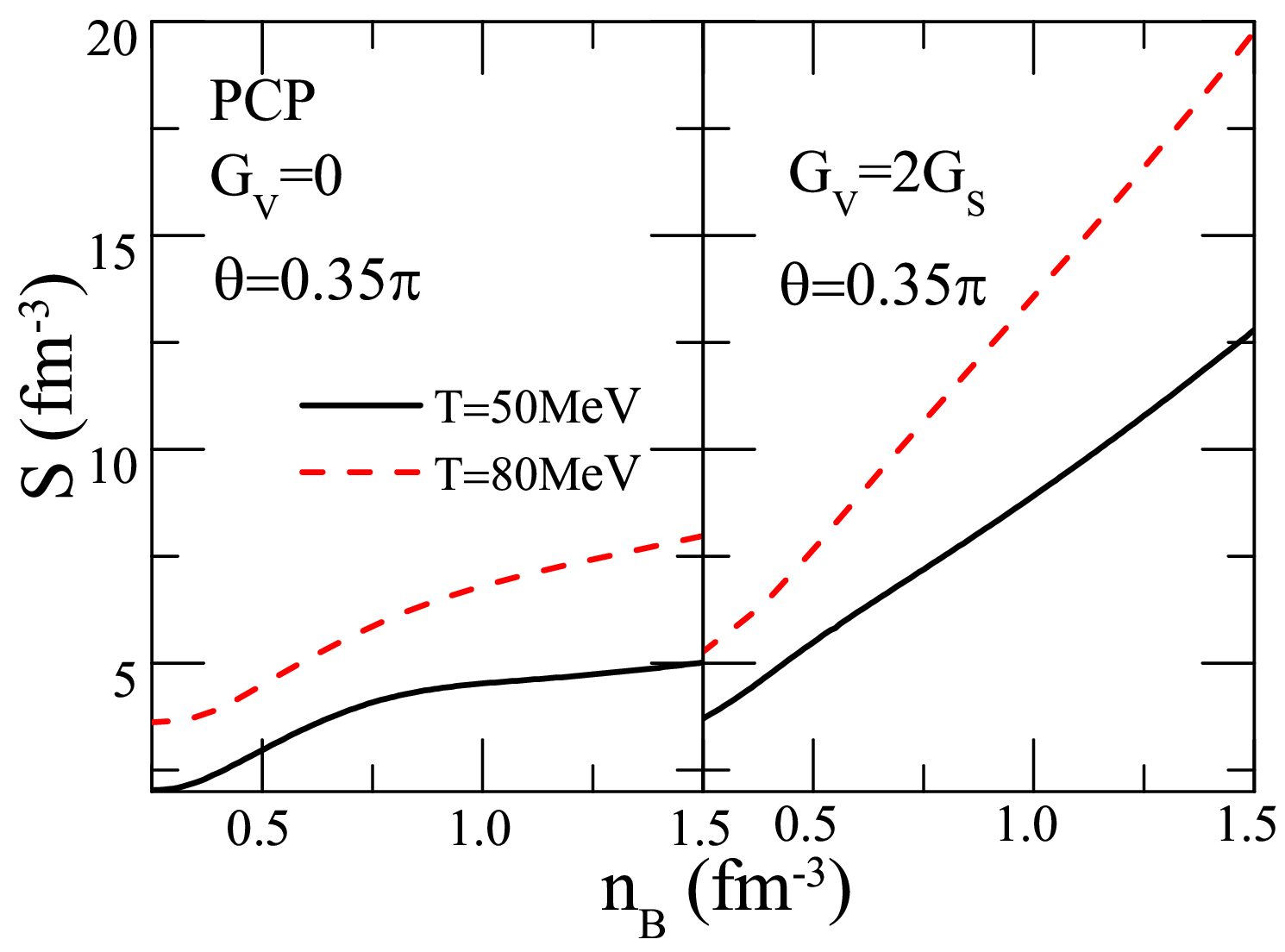}
\caption{(Color online) The entropy density as a function of $n_B$ at finite temperature with different $G_V$. }
\label{EOS}
\end{figure}

In Fig. 4, we calculate the entropy density $S$ as a function of $n_B$ at finite temperatures with varying $G_V$ when $\theta=0.35\pi$. It is evident that for all cases, the entropy density increases monotonically with the baryon density at finite temperature. Furthermore, the entropy density is significantly enhanced by a larger vector coupling constant $G_V$ and higher temperature. Our detailed calculations also reveal that the dependence of the entropy density on $\theta$ is weak, exhibiting only a slight decrease as $\theta$ increases.

\begin{figure}[tbp]
\includegraphics[scale=0.35]{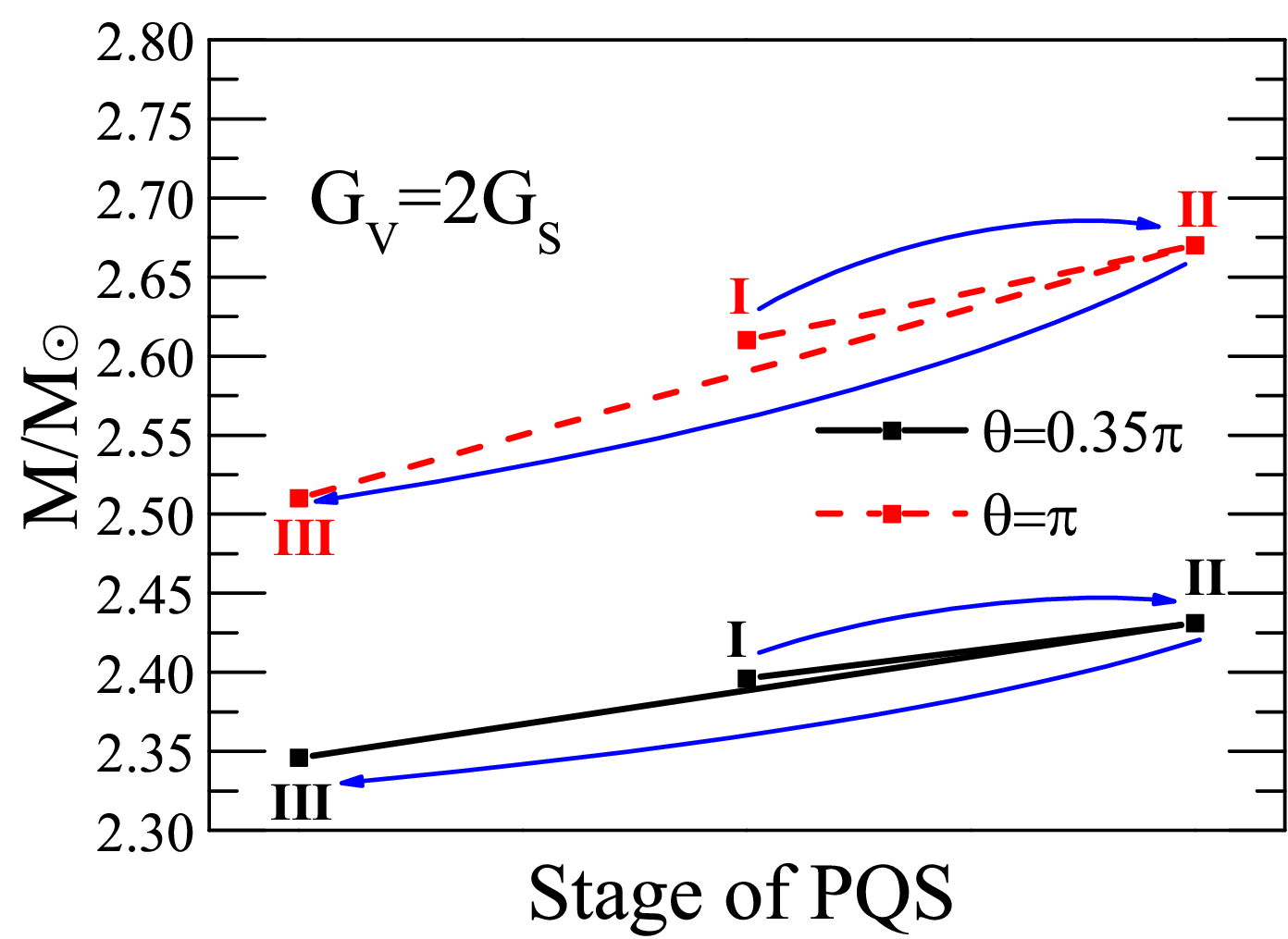}
\caption{(Color online) The maximum star mass at different stages of the proto-quark star with different $\theta$.}
\label{RM}
\end{figure}

As shown in Fig. 5, we calculate the maximum mass of PQSs at different isentropic stages along the star evolution line with different values of $\theta$. It can be seen that the maximum star mass of PQSs increases from the 1st stage ($S/n_{B}=1,~Y_{l}=0.4$) to the 2nd stage ($S/n_{B}=2,~Y_{\nu_{l}}=0$). This suggests that the temperature effects and the transition involving neutrino diffusion significantly stiffens the equation of state, thereby supporting a larger stellar mass. However, at the 3rd stage, where the neutrino fraction drops to zero and the star cools down to zero temperature, the maximum star mass decreases. Specially, for $\theta=\pi$ case, the maximum star mass of PQSs rises from $2.61~M_{\odot}$ at the 1st stage to a peak of $2.67~M_{\odot}$ at the 2nd stage, while decreasing to $2.51~M_{\odot}$ at the 3rd stage (the ordinary zero temperature compact star case). Furthermore, Fig. 5 demonstrates that for any fixed evolutionary stage, the maximum star mass increases as $\theta$ varies from $0.35\pi$ to $\pi$. This indicates that axion-induced effects within the SU(3) NJL model enhance the stiffness of the EoS of the star matter at finite temperature (which is consistent with the conclusion from Fig. 2 ), allowing for more massive compact stars.

\begin{figure}[tbp]
\includegraphics[scale=0.34]{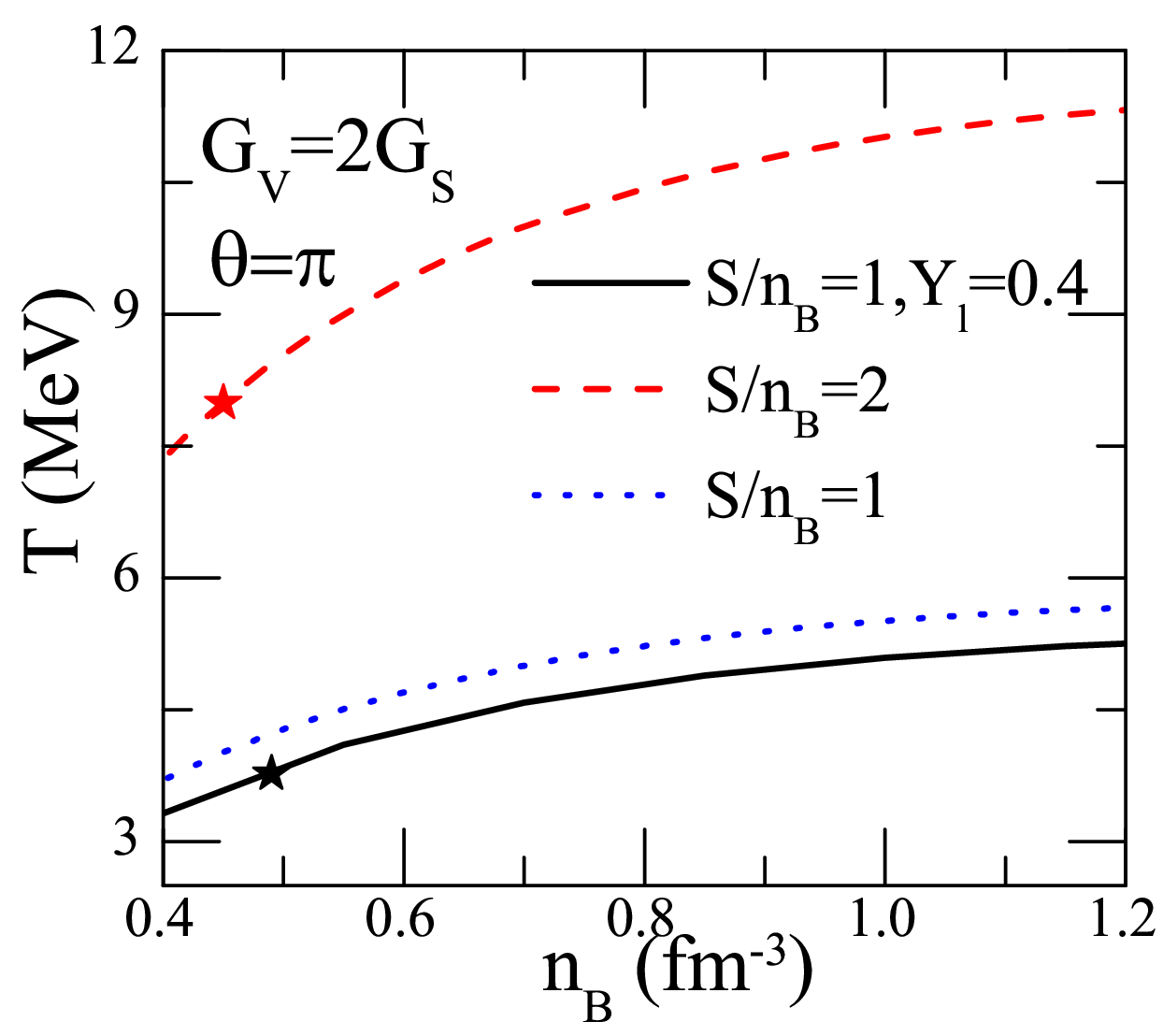}
\caption{(Color online) The temperature of the proto-quark star matter with different isentropic stages along the star evolution.}
\label{RM}
\end{figure}

In Fig. 6, we present the temperature of proto-quark star matter at different isentropic stages along the star evolution line.
It can be seen that the temperature of the star matter increases with baryon density for all cases. Notably, at a fixed baryon density $n_B$, the temperature of the 2nd stage ($S/n_{B}=2,~Y_{\nu_{l}}=0$) is significantly higher than that at the 1st stage ($S/n_{B}=1,~Y_{l}=0.4$). Furthermore, we include a comparison case represented by the blue dotted line with $S/n_{B}=1,~Y_{\nu_{l}}=0$ in Fig. 6. One can observe that removing the trapped neutrinos leads to a higher temperature compared to the 1st stage. This behavior arises because both temperature and the number of particle species contribute to the entropy density of quark matter. The presence of the trapped neutrinos in the 1st stage $S/n_{B}=1,~Y_{l}=0.4$ provides an additional contribution to the entropy. Consequently, a lower temperature is sufficient to maintain the fixed entropy per baryon $S/n_{B}=1,~Y_{\nu_{l}}=0$, which explains why the 1st stage exhibits a lower temperature than the neutrino-free case (blue dotted line).
Moreover, we also calculate the core temperature corresponding to the maximum star mass of PQS (indicated by the star symbols on the curves). The results show that the core temperature increases from the 1st to the 2nd stage, while the corresponding central baryon density decreases.

\begin{figure}[tbp]
\includegraphics[scale=0.36]{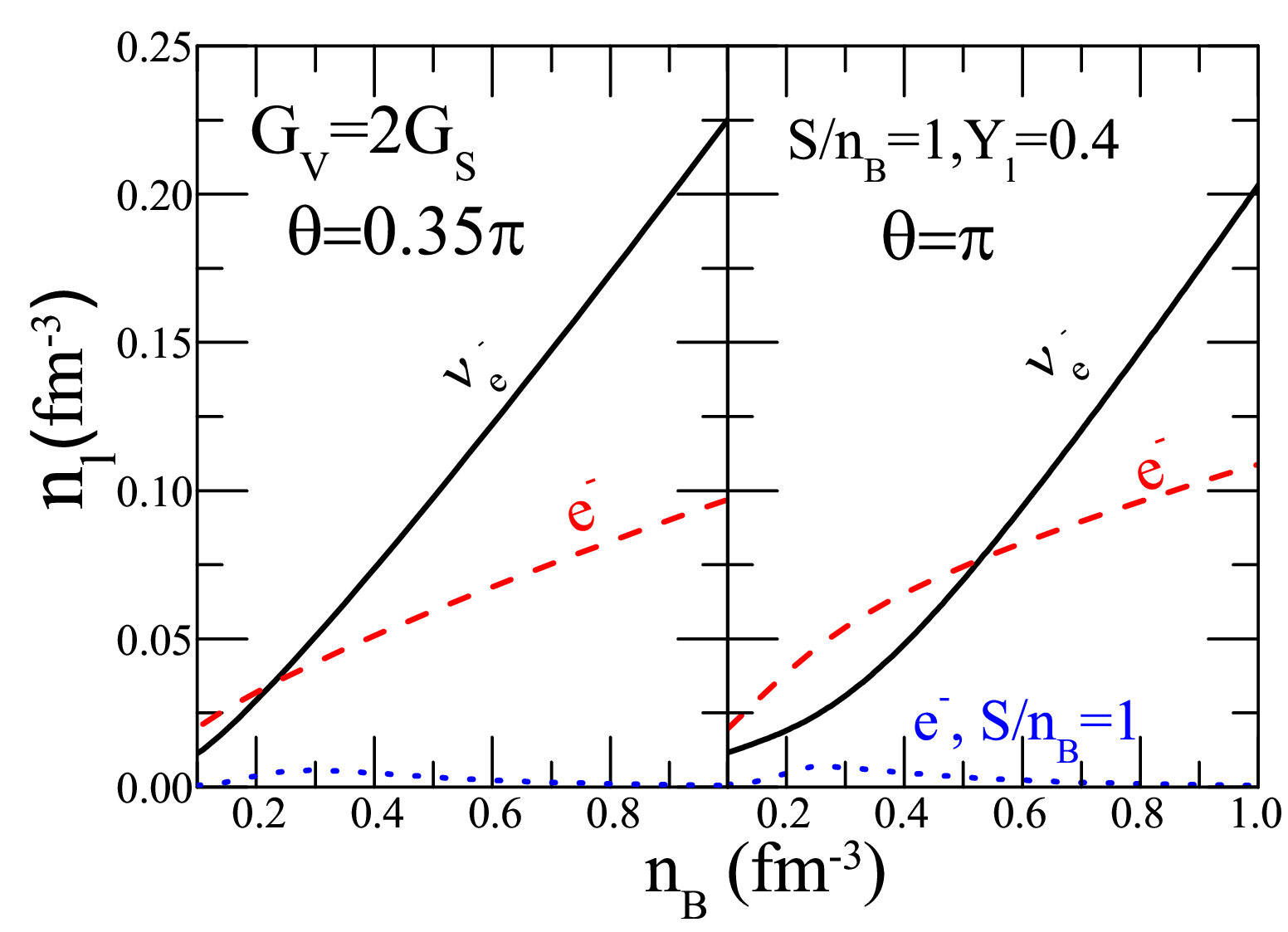}
\caption{(Color online) The number densities of neutrinos and electrons as functions of $n_B$ when $S/n_B=1,~Y_l=0.4$ and $S/n_B=1,~Y_{\nu_{l}}=0$.}
\label{RM}
\end{figure}
\begin{figure}[tbp]
\includegraphics[scale=0.36]{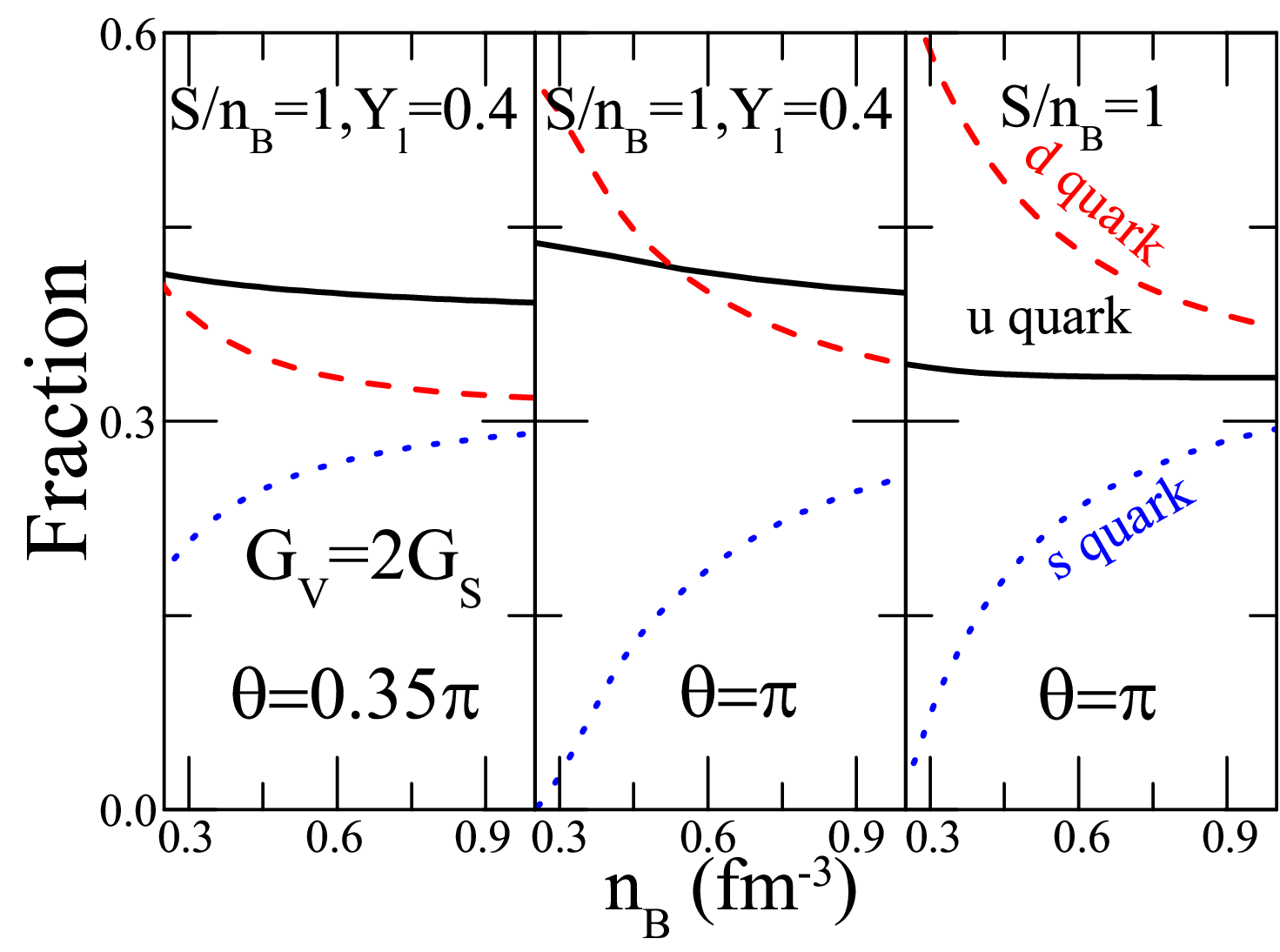}
\caption{(Color online) Fractions of $u$, $d$ and $s$ quarks with different cases at finite temperature.}
\label{RM}
\end{figure}
In Fig. 7, we calculate the number densities of neutrino and electron as functions of $n_B$ with different $\theta$ when $S/n_B=1,~Y_l=0.4$ and $S/n_B=1,~Y_{\nu_{l}}=0$. It can be seen in Fig. 7 that the number densities of electron increases with $\theta$ while the number densities of neutrino decreasing when $S/n_B=1,~Y_l=0.4$. Furthermore, the number densities of electron in $S/n_B=1,~Y_{\nu_{l}}=0$ case significantly becomes smaller than that of $S/n_B=1,~Y_l=0.4$ when $\theta=\pi$.

In Fig. 8, we further calculate the fractions of $u$, $d$ and $s$ quarks as functions of $n_B$ with different $\theta$ when $S/n_B=1,~Y_l=0.4$ and $S/n_B=1,~Y_{\nu_{l}}=0$. As observed in the left and middle panels of Fig. 8, for the case with trapped neutrinos $S/n_B=1,~Y_l=0.4$, the fraction of $d$ quark increases with $\theta$, while that of $s$ quarks decreases. Meanwhile, the fraction of $u$ quark increases with $\theta$ to satisfy charge neutrality. In the right panel, the sum of the fractions of $d$ and $s$ quarks is approximately twice that of the $u$ quarks. This explains the reason, as shown in Fig. 7, why the number densities of electron in $S/n_B=1,~Y_{\nu_{l}}=0$ is significantly lower than that in the trapped neutrino case ($S/n_B=1,~Y_l=0.4$) when $\theta=\pi$.

\section{conclusion and discussion}
In this work, we investigate the thermodynamical properties of SQM and PQSs within the SU(3) NJL model at finite temperature, specifically incorporating the effects of axion fields and vector interactions.

We first examine the minimum (free) energy per baryon and the pressure as functions of the free energy density at various temperatures by considering the axion effects and vector interactions in SU(3) NJL model. From the calculations, we find that the stability of SQM is greatly enhanced by the axion field effect in the finite temperature regime, and the EoS of SQM becomes stiffer with increasing axion effects and temperature.

Furthermore, we calculate the constituent quark mass of $u$ and $d$ quarks and the entropy density at finite temperature with different $\theta$. The results indicate that the constituent quark mass decreases with the axion effects at finite temperature,with the mass of $d$ quark being smaller than that of $u$ quark, while the entropy density is significantly enhanced by the vector interactions and the temperature effect. The detailed calculations also reveal that the dependence of the entropy density on $\theta$ is weak, exhibiting only a slight decrease as $\theta$ increases.

Moreover, we study the maximum mass of PQSs at the isentropic stages along the star evolution line, and the results indicate that axion effects and temperature effects can enhance the stiffness of the EoS of the star matter at finite temperature and further supporting more massive compact stars. We then investigate the temperature of the proto-quark star matter and the lepton number density of the isentropic heating stage, and we find the core temperature of the maximum star mass increases with the entropy per baryon while the corresponding central baryon density decreasing. Notably, we also find the number density of electrons increases in the presence of trapped neutrinos, whereas the star matter temperature decreases compared to the neutrino-free case.

Therefore, our results demonstrate that including axion effects and the vector interactions in SU(3) NJL model at finite temperature can significantly influence the equation of state, constituent quark mass, entropy density, and the maximum star mass of PQSs at the isentropic stages along the star evolution line. In particular, the presence of trapped neutrinos leads to a substantial increase in electron number density while simultaneously suppressing the core temperature. These findings are crucial for understanding the observable properties of hot compact stars.

\vskip 1 cm
{\bf Acknowledgement.---} This work is supported by the NSFC under
Grants No. 12575134, 12505159, 11975132, 12205158, 12305148, and
11505100, 12575137 the Shandong Provincial Natural Science
Foundation, China ZR2022JQ04, ZR2025QC1487, ZR2021QA037, ZR2019YQ01 and ZR2015AQ007, and the Natural Science Foundation of Qingdao, China (Grant No. project 25-1-1-4-zyyd-jch).

\end{document}